# RESPONSE OF INITIAL FIELD TO STIFFNESS PERTURBATION

Chen-Wu Wu *
Institute of Mechanics, Chinese Academy of Sciences, No.15 Beisihuan Xi Road, Beijing 100190, China.
E-mail: chenwuwu@imech.ac.cn & wuchenwu@gmail.com

**ABSTRACT**

Response of initial elastic field to stiffness perturbation and its possible application is investigated. Virtual thermal softening is used to produce the stiffness reduction for demonstration. It is interpreted that the redistribution of the initial strain will be developed by the non-uniform temperature elevation, as which leads to the non-uniform reduction of the material stiffness. Therefore, the initial filed is related to the stiffness perturbation and incremental field in a matrix form after eliminating the thermal expansion effect.

## 1. INTRODUCTION

Under the action of external loading or internal eigen strain, the elastic field of the solid is determined by the material stiffness. If the material stiffness undergoes a perturbation, like induced by localized temperature, the elastic field should be redistributed. An observable incremental field would appear during such redistribution. Obviously, this incremental field is determined by the initial field, the stiffness characteristics of the solid before and after perturbation. Once the initial field can be backward deprived through the incremental field and the stiffness characteristics, one can use such a stiffness perturbation method to analyze the initial field.

There is a kind of eigen strain problem, which is always known as so called residual stress, of which the measurement is of crucial for the performance evaluation on the materials and structures. Generally speaking, aside from some subtle methods (like Thermo-elastic Stress Analysis: TSA), the common methods for evaluating the residual stresses can be divided into two categories, i.e., mechanical and physical methods. The mechanical methods measure the released strain after relaxing the residual stress through some mechanical means (Schajer, 2010; Withers 2001a), such as the method of hole-drilling, splitting, sectioning and layer removing etc., which are generally considered to be destructive. Actually, according to the view of the present article, almost all of the mechanical relaxation methods can be thought of as one kind of stiffness perturbation, with the stiffness within the cut region being reduced to zero. The physical methods measure directly the lattice constant deviation due to residual stress through some diffraction test (Withers 2001a), such as electron diffraction, x-ray diffraction and neutron diffraction etc.

The concept of temporarily thermal relaxation (Wu, 2012) is developed as an alternative to make a stiffness perturbation thereafter to assess the residual stress in film (Cheng, 2013). According to Hooke' law, the internal stress in a self-equilibrium solid is accompanied with elastic strain (Withers 2001b), of which the magnitude is correlated to the stiffness of the material. Moderate temperature elevation of the solid will lead to the reduction of its stiffness, i.e., the elastic modulus E (Fernandes, 1973; Pitarresi G, 2003). Such non-uniform thermal softening will lead to the redistribution of the initial strain field, which will contribute a displacement increment to the total observable displacement in the initially stressed solid subjected to thermal loading. This article is focused on separating the contribution of the temporarily thermal relaxation of the initial (residual) stress to the total observable displacement.

## 2. THEORETICAL INTERPRETATION

Consider a solid domain, within which some eigen strain field arises during manufacturing, thus there is a residual stress field $\boldsymbol{\sigma}_0$ accompanied with the initial eigen strain field as shown in Fig.6 (a). Of course, such residual stress/ strain should develop a displacement field of the material points, although we could not observe it just through its post-manufacturing state. According to the theory of elastic solid, there exist relevant equations controlling the deformation and stress distribution.

First of all, the equations of equilibrium (Landau LD, 1986) for a static/ quasi-static solid can be expressed as

$$\sigma_{ij,j} = 0, \ i = 1, 2, 3 \quad (1)$$

where $\boldsymbol{\sigma}$ is the two order stress tensor.

Secondly, the Hooke's law for the isotropic solid of temperature-dependent elastic parameters is

$$\sigma_{ij} = 2\mu(\theta)\gamma_{ij} + \lambda(\theta)\gamma_{kk}\delta_{ij} - \beta(\theta)\theta\delta_{ij}, \ i,j = 1, 2, 3, \quad (2)$$

where $\mu(\theta)$ and $\lambda(\theta)$ are temperature-dependent Lame constants, $\theta$ represents temperature, $\boldsymbol{\gamma}$ is the two order strain tensor, $\beta(\theta) = (3\lambda(\theta) + 2\mu(\theta)) \times \int_0^\theta \alpha(T)dT$ is the temperature-dependent thermo-mechanical coefficient with $\alpha(\theta)$ being the thermal expansion coefficient and $\delta_{ij}$ represents the Kronecker' delta.

Moreover, the components of the strain tensor $\boldsymbol{\gamma}$ can be related to the displacement vector $\mathbf{u}$ with

$$\gamma_{ij} = (1/2)(u_{i,j} + u_{j,i}), \ i,j = 1, 2, 3. \quad (3)$$

Now, consider the initial state with only the self-equilibrium residual stress being acted on the solid, of which the temperature is uniform and identical to the ambient temperature. This indicates that the Lame constants will be identical everywhere within the whole solid and no thermal expansion arises. So the Hooke's law (2) reduces to

$$\sigma_{ij} = 2\mu\gamma_{ij} + \lambda\gamma_{kk}\delta_{ij}, \ i,j = 1, 2, 3, \quad (4)$$

Therefore, one can substitute (3) and (4) into (1) and obtain

$$\mu\Delta u_i + (\lambda + \mu)e_{,i} = 0, \ i = 1, 2, 3 \quad (5)$$

with the operator

$$\Delta = \partial^2/\partial x_1^2 + \partial^2/\partial x_2^2 + \partial^2/\partial x_3^2, \quad (6)$$

and the volume strain

$$e = u_{i,i}, \ i = 1, 2, 3. \quad (7)$$

For a given initially state of self-equilibrium residual stress, there is a displacement $\mathbf{u}_0$ field relative to the stress free state.

Once the solid is partially heated and a non-uniform temperature field arises within it, the temperature of every material point can be expressed as function of the space coordinates

$$\theta = \theta(\mathbf{x}). \quad (8)$$

Therefore, the temperature-dependent Lame constants and thermo-mechanical coefficients of every material point will depend actually on its space position, that is

$$\mu(\theta) = \mu(\theta(\mathbf{x})) = \mu(\mathbf{x}), \quad (9)$$
$$\lambda(\theta) = \lambda(\theta(\mathbf{x})) = \lambda(\mathbf{x}), \quad (10)$$
and
$$\beta(\theta) = \beta(\theta(\mathbf{x})) = \beta(\mathbf{x}), \quad (11)$$

Thus, one can again substitute (2) and (3) into (1) and obtain

$$\mu(\mathbf{x})\Delta u_i + (\lambda(\mathbf{x}) + \mu(\mathbf{x}))e_{,i} + \mu_{,j}(u_{i,k} + u_{k,i}) + \lambda_{,i}u_{k,k} - \beta(\mathbf{x})\theta_{,i} - \beta_{,i}\theta = 0, \quad i = 1,2,3. \quad (12)$$

In other words, the initial state (as sketched in Fig.1 (a)) is determined by the eigen strain and the elastic constants of that state, while the present state (as sketched in Fig.6 (b)) is determined both by the eigen strain, the present elastic constants and the non-uniform thermal expansion. By assuming that the thermal expansion coefficient is zero all along, one can understand that the displacement field $\mathbf{u}_0$ developed by the initial residual stress will be changed by the non-uniform drifting of the Lame constants due to the non-uniform temperature elevation. Because, different stress and deformation will be developed by the same eigen strain for the two states of different elastic constants.

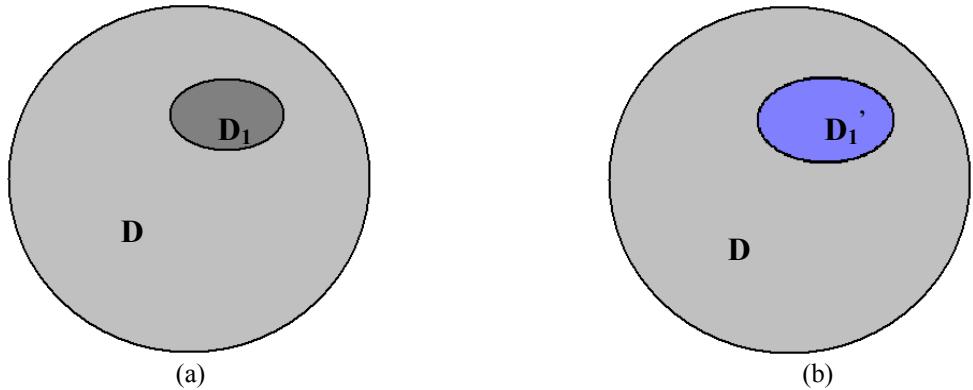

(a)          (b)

Fig.6 Sketch of (a) the initially strained solid and (b) the redistribution of the initial displacement

That is, after the non-uniform temperature elevation is introduced into the structure by partial heating, the displacement relative to the virtual stress-free state would be

$$\mathbf{u} = \mathbf{u}_0 + \mathbf{u}'. \quad (13)$$

Where, the item $\mathbf{u}'$ represents the displacement field of the present state relative to the initial state of the solid under the only action of the self-equilibrium residual stresses. Obviously, it is observable in the thermal loading experiment, therefore can be measured and analyzed by many techniques, like combining the Digital Image Correlation method and thermo-elastic analysis. Furthermore, this observable displacement vector $\mathbf{u}'$ is composed of two parts. Namely,

$$\mathbf{u}' = \mathbf{u}_{TE} + \mathbf{u}_0^r. \quad (14)$$

In the right side of equation (14), the first part $\mathbf{u}_{TE}$ is developed by the non-uniform thermal expansion and the second part $\mathbf{u}_0^r$ is induced by the thermal relaxation of the initial residual stress, which is determined both by the initial displacement vector $\mathbf{u}_0$ and the elastic constants, i.e. $\mu(\theta)$ and $\lambda(\theta)$, of both states of ambient temperature and elevated temperature.

If the domain is discretized into finite elements, one have the equilibrium equation as for the structure under the action of initial stress, namely

$$[K_0][u_0] = \{P\}. \quad (15)$$

Once the stiffness matrix is altered slightly by the non-uniform temperature elevation while no thermal expansion is taken into account, the equilibrium should be replaced by

$[K_1][u_1] = \{P\}$.  (16)

Let

$u_0 + \Delta u = u_1$,  (17)

and rewrite (16) as

$[K_1][u_0] + [K_1][\Delta u] = \{P\}$.  (18)

Combining (18) and (15), one can obtain

$([K_1] - [K_0])[u_0] + [K_1][\Delta u] = 0$.  (19)

$[u_0] = [K_1][\Delta u] / ([K_0] - [K_1])$.  (20)

Let

$[\Delta K] = [K_1] - [K_0]$,  (21)

Equation (20) can be reduced as

$[u_0] = -[K_1][\Delta u] / [\Delta K]$.  (22)

Or rather,

$[u_0] = -([K_0] + [\Delta K])[\Delta u] / [\Delta K]$.  (23)

That is,

$[u_0] = -([K_0]/[\Delta K] + 1)[\Delta u]$.  (24)

It is indicated that the initial elastic field would be determined by the initial stiffness, stiffness perturbation and the incremental field.

Further, were the stiffness perturbation small enough, therefore the displacement increment always should be very small, then one can gat that

$[u_0] \approx -[K_0] \dfrac{[\Delta u]}{[\Delta K]}$.  (25)

## 3. DISCUSSION

The non-uniform thermal softening of the material would lead to partial relaxation of the pre-existed stress in an initially stressed solid subject to non-uniform heating. The total observable displacements arise during the thermal loading experiment is determined by the initial (residual) stress, the thermal softening effect and the thermal expansion effect. The incremental displacements rely on the partially released residual stress and the temperature-dependent elastic constants. Therefore, such concept of temporarily thermal relaxation may be an alternative for residual stress assessment once the relationship between the displacement increment and the initial stress is established, which should be carried through according to the characteristics of the object under consideration.

## ACKNOWLEDGMENTS

The author gratefully acknowledges the funding by National Natural Science Foundation of China, under grants No. 11002145.